# Handling Numerous Stakeholders' Requirements - A Research Agenda and Preliminary Outcomes

Saurabh Malgaonkar, Sherlock A. Licorish, Bastin Tony Roy Savarimuthu
Department of Information Science
University of Otago
Dunedin, New Zealand
malsa876@student.otago.ac.nz, sherlock.licorish@otago.ac.nz, tony.savarimuthu@otago.ac.nz

## ABSTRACT
This research aims to design and develop a new requirements prioritization approach for analyzing and prioritizing stakeholders' requirements which are mentioned in the feedback for software products. This paper presents a PhD research agenda and preliminary outcomes from early analysis. A roadmap to the proposed research methodology that is to be followed to achieve the targeted outcomes is also outlined. Outcomes to date show that the requirements prioritization problem has been researched extensively, however, gaps still remain when considering techniques that handle a large number of crowdsourced requirements. Furthermore, requirements prioritization as a problem affects many domains beyond software engineering. Hence, knowledge from other fields could be useful for informing requirements prioritization practice in the software engineering space.

## 1 INTRODUCTION
The prime purpose of a software product is to satisfy its stakeholders' needs or demands. These needs and demands come in the form of 'requirements'. The requirements engineering phase, and subsequent phases of the software development process, transform these requirements into a fully functional product. These requirements typically get captured by the software team before the actual design and development of the software product. Additionally, when the software product is launched, the requirements come in the form of user feedback. This feedback contains useful information such as suggestions to make the software product better, requests for additional features, and particular problems encountered while using the software product [1]. This research project deals with the prioritization of users' requirements. The primary goal of our research is to identify and extract the requirements of stakeholders, and prioritize these so that they can be suitably addressed.

Currently, in the mobile app market, there are millions of applications available for the most commonly used Android and Apple mobile devices [2]. The developers of these apps are continuously improving the features or fixing the commonly faced problems in view of helping end-users to have a positive experience. The developers find these elements in the submitted user feedback, and must optimize the order in which problems are fixed [3]. User feedback, unlike formal documentation, allows users to express their opinion on the app features freely, and in fact, such opinions are integral for the improvement of the quality of services offered and enhancement that could be made to software deliverables. Feedback on a software product provides a good understanding of the reception level of a software product among users. While the utility of reviews for developers and customers is well-known [4], these reviews are enormous, and the challenge to extract requirements from such crowdsourced data for prioritization purposes still exits [5]. Traditional requirements prioritization techniques are not able to meet this challenge [6].

One of the problems with voluminous app review data is that the clients' data on app reviews are stored in vast storage networks. Apart from timely access to such massive data, other problems such as ambiguity and inconsistent spellings need to be addressed [7]. Such data cannot be discarded and need to be cleaned to obtain better results [8]. Considering all these facts, we plan to develop a solution that handles and prioritizes requirements while operating in a crowdsourced environment. In addressing the challenge of prioritizing requirements, we intend to undertake the following three activities:

1. Perform data pre-processing and conduct experiments around variances in text/review semantics for extracting only the precise requirements of end users amidst irrelevant data.
2. Identify and evaluate the requirements in reviews with reliable predictive power. This predictive power will help to automatically rank the requirements based on their severity and importance.
3. Evaluate suitable intelligent techniques for ranking candidate features which reflect the requirements of the app users.

In this paper, we highlight the overview of our proposed research and the relevant tasks completed to date. In Section 2 we highlight the research objectives of our project, and the core activities that have been planned. This section also identifies the research gaps. Section 3 provides the completed work to date, and Section 4 summarizes this paper and describes the next phases of our work.

## 2 OVERVIEW OF RESEARCH
This section describes the background of the proposed research, the identified research gaps, the primary research objectives of the proposed research, and the formulated work plan to complete this research.

### 2.1 Background
Requirements prioritization deals with the ranking or classification of user requirements based on their severity and importance [9]. Identification of prioritized requirements assists developers with releasing software products with user-solicited functionalities and launching necessary software updates. This is particularly critical when there are numerous requirements in the form of user feedback. Our research primarily targets the 'feedback' that is provided after a software product is released. This feedback contains information regarding the requirements that should be addressed. We plan to investigate the feedback which is provided in relation to mobile applications by end-users [10, 11]. The goal of the study is to identify and extract requirements mentioned in the feedback and prioritize these requirements according to users' preferences. This is a very challenging task, as we were able to identify a crucial research gap that still exists in the literature despite many years of research. This issue is considered further in the next section.

### 2.2 Research Gap
Keertipati et al. [12] have worked on extracting the requirements of customers that are mentioned in mobile app reviews, and have implemented multiple techniques that prioritize those extracted requirements. Later, Licorish et al. [13] took a step back and carefully studied features that are evident is user reviews that may predict their urgency. They mentioned that the challenge related to handling and prioritizing numerous requirements for mobile app reviews still exist. Similar problem related to handling and prioritizing large amounts of requirements was highlighted in the study carried out by Karlsson et al. [14] for market-driven software products. In a crowd-sourced context, requirements always tend to be enormous; leading to the next release problem [15]. The problem states that software teams always face the challenge of ranking requirements that need to be addressed first, as it is not possible for the team to fix all the requirements in one go. This is predominantly due to time, budget or resource constraints that are imposed on the software team. This is where the application of requirements prioritization is essential. The requirements prioritization process helps to rank the requirements, which enables the software team to address the requirements in multiple stages. Using this approach the software team can pinpoint and fix the urgent requirements in the early stages, and the least important ones could be fixed later. In a study carried out by Laurent et al. [16], the importance of prioritizing a large number of requirements automatically with less human intervention is emphasized, and the study noted that the prioritization techniques that follow such an approach are scarce, and in most cases are custom design methods. This work thus clearly highlights the research gap in this domain, and this research aims to bridge this gap.

### 2.3 Research Objectives
The primary objective of this study is to address the issue of handling numerous requirements by developing a novel requirements prioritization technique. In order to achieve this, the following activities are scheduled:

Phase 1: Perform a systematic mapping study [17] of the requirements prioritization domain to understand what has been done on this topic across all disciplines.

Phase 2: Filter out the top quality literature from the shortlisted articles based on the outcome of the systematic mapping study and evaluate the solutions that provide the most value, and particularly those research works where the authors have presented an evaluated solution.

Phase 3: Develop a solution that has the strengths of the best evaluated solutions, and overcomes their drawbacks.

Phase 4: Apply the new solution to a real-world problem. The targeted domain is 'Mobile App Reviews'.

Mobile app reviews contain information regarding the requirements that need to be fulfilled from the users' perspective. These requirements need to be identified and prioritized so that they could be fixed at the earliest, which will help the mobile app in having a competitive edge in the software market. The primary objectives of our proposed requirements prioritization technique will be: to handle and process a large number of reviews from numerous app users, automating the ranking process of reviews and providing practical solutions for setting up clear priorities.

### 2.4 Research Work Plan
This PhD work is planned as follows:
• Year 1: Evaluating the various requirements prioritization techniques and creating a taxonomy for the comparison of these techniques (refer to Phase 1 in Section 2.3).
• Year 2: Designing and developing a requirements prioritization technique that addresses the drawbacks of the existing evaluated techniques and combines their strengths. This aspect of the work will also compare the performance of the known techniques against the newly developed technique (refer to Phases 2 and 3 in Section 2.3).
• Year 3: Testing the developed technique at the industry level and writing up the PhD thesis (refer to Phase 4 in Section 2.3)

## 3 PROGRESS TO DATE
We have initiated a systematic mapping study towards understanding the works that have focused on requirements prioritization. This section describes the application of this method in our research and provides preliminary results aimed at answering our first research question (included below).

### 3.1 Systematic Mapping Study
The systematic mapping study helped us to discover the relevant literature on requirements prioritization from a broad perspective, but at the same time assisted us in selecting good quality articles for further review. Using the systematic mapping study, we were able to define the problem statement on requirements prioritization and formulate the research questions for Phase I of the research.

Using specific evaluation criteria we were able to shortlist articles (211 altogether) from 5,225 articles that appeared in the search results from eight databases (IEEE Xplore, ACM, Springer, Scopus, Web of Science, ScienceDirect, Inspec and EI Compendex). The main advantage of the systematic mapping study is that it allows the authors to classify and visualize the research available on a particular topic for better understanding [18]. We are currently in the stage of finalizing the systematic mapping study of the requirements prioritization domain, and are documenting related works. For Phase 1 of the research indicated in Section 2.3, we have framed five research questions. The framed research questions are as follows:

RQ1. What has been the interest in requirements prioritization over time, what are the different publication venues and what are the various domains of their application?

RQ2. What approaches have been used to study requirements prioritization?

RQ3. What form did the contribution of the requirements prioritization techniques take?

RQ4. What prioritization techniques have been studied or developed?

RQ5. What research methods are followed by those investigating requirements prioritization techniques?

These questions allow us to carry out an in-depth examination of the requirements prioritization domain and cover the relevant studies residing in this domain comprehensively. RQ1 helps to uncover the level of interest of the researchers in the requirements prioritization domain, and also identifies the various publication platforms targeted by the researchers to publish their work on requirements prioritization. RQ2 assists in identifying the nature of research which has been carried out by the researchers, while RQ3 indicates the type of contribution made by the researchers. The various requirements prioritization techniques that are available to prioritize the requirements are scrutinized by answering RQ4. Finally, RQ5 identifies the various data collection and data analysis approaches that are followed by the researchers to carry out their research on requirements prioritization.

By means of the systematic mapping study, we were able to design the research questions and conduct searches in the appropriate knowledge databases. We are currently performing the analysis of the outcomes of this study which also include the identification of various requirements prioritization techniques. Later, we will provide a summary of results of the systematic mapping study and develop a report on it.

We provide preliminary outcomes to answer the first question of the 'Phase 1' of our research in the next section.

## 3.2 Interest In Requirements Prioritization, Publication Venues And Domains Of Application

We answer the first research question to provide insights on the level of interest of researchers in the field of requirements prioritization over the past years, along with the various venues that are utilized by the researchers to publish their work on this topic. We were also able to find out the application of requirements prioritization across various domains.

Figure 1 demonstrates the level of interest of researchers in the requirements prioritization domain where it is noted that prior to 2004 the number of publications were few. However, since 2004 the number of publications have increased, indicating that there is growing interest in the topic among researchers.

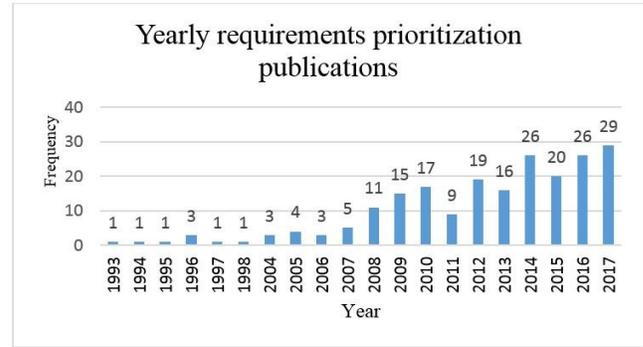

**Figure 1. Yearly requirements prioritization publications**

Table 1 shows that conferences and journals were targeted most by researchers to publish their work on requirements prioritization, while very few studies were published as book chapters. Some of the studies were delivered through workshops and symposiums, and only one study was presented at a world forum.

**Table 1. Requirements prioritization publication venues**

| Venue | Total |
|---|---|
| Conference | 103 |
| Journal | 75 |
| Workshop | 16 |
| Symposium | 7 |
| Chapters | 9 |
| World Forum | 1 |

Table 2 indicates that the application of requirements prioritization is most common in the field of software engineering, followed by product manufacturing. However, we also discovered that requirements prioritization was also studied in other domains, including real estate, education, finance, product manufacturing, transport, and law.

**Table 2. Requirements prioritization publication domains**

| Domain | Total |
|---|---|
| Software Engineering | 175 |
| Real Estate | 3 |
| Finance | 4 |
| Product Manufacturing | 22 |
| Education | 5 |
| Transport | 1 |
| Law | 1 |

Figure 2 visualizes the venue and domain results. Here it can be observed that the majority of the articles published in the software engineering domain were published in conferences and journals, with conferences tending to dominate. On the other hand, for other venues (e.g., product manufacturing) researchers seem to favor journals for publication. Interestingly, Figure 2 shows that requirements prioritization was also discussed at a world forum as part of product manufacturing.

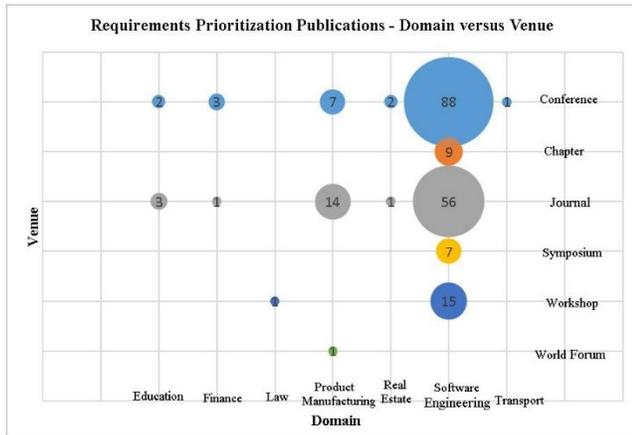

**Figure 2. Requirements prioritization publications - domain versus venue**

## 4 SUMMARY AND FUTURE WORK

The systematic mapping study helped us to retrieve literature on requirements prioritization from all domains (beyond Software Engineering). This will assist us in designing our requirements prioritization framework by taking inspirations from the requirements prioritization techniques which have their application across various domains. It is our opinion that the systematic mapping study guides research on a topic in the right direction, which also aids in undertaking a detailed review of the literature. We are in the process of addressing the remaining questions of our systematic mapping study, and will subsequently document the results for the research questions. This will lead to the completion of phase one of the research study. After phase one is completed, we will proceed to the second phase which will be the detailed evaluation of the various requirements prioritization techniques. This will be achieved by conducting a full detailed literature review of the articles that are shortlisted as a result of the systematic mapping study. We will also be investigating the type of data that we will be dealing with (mostly textual) while designing our requirements prioritization model. The experimentation platform of our system will help us to evaluate various requirements prioritization techniques against our proposed requirements prioritization technique. Thereafter, we will document our inferences, conclusions, and future directions.

We anticipate that mobile app reviews will provide our testbed once we have formulated the necessary stages of our requirements prioritization technique. To this end, there will be a need for the extraction of essential requirements that are submitted by mobile app users. These requirements will then be prioritized on the basis of their importance and severity. We will then tune the performance of the requirements prioritization technique concerning the accuracy of prioritization (ranking), time utilization, scalability, and so on. The design and development of this proposed model is however subject to change depending upon the progress of our study and findings of our work. The application of this research will be useful for the qualitative feedback analysis of mobile apps or in gauging public opinions on mobile apps and building support bases for mobile app developers.

## ACKNOWLEDGMENTS

Phase 1 of this study is funded by a University of Otago Research Grant.

## REFERENCES

1. Sommervile, I., *Software Engineering*. 9 ed. 2009.
2. Saifi, R. cited 2017; Available from: https://www.business2community.com/mobile-apps/2017-mobile-app-market-statistics-trends-analysis-01750346.
3. Fu, B., et al., *Why people hate your app: making sense of user feedback in a mobile app store*, in *Proceedings of the 19th ACM SIGKDD international conference on Knowledge discovery and data mining*. 2013, ACM: Chicago, Illinois, USA. p. 1276-1284.
4. Su'a, T., et al., *QuickReview: A Novel Data-Driven Mobile User Interface for Reporting Problematic App Features*, in *Proceedings of the 22nd International Conference on Intelligent User Interfaces*. 2017, ACM: Limassol, Cyprus. p. 517-522.
5. Hosseini, M., et al. *Configuring crowdsourcing for requirements elicitation*. in *2015 IEEE 9th International Conference on Research Challenges in Information Science (RCIS)*. 2015.
6. Achimugu, P., et al., *A systematic literature review of software requirements prioritization research*. Information and Software Technology, 2014. **56**(6): p. 568-585.
7. Iacob, C. and R. Harrison. *Retrieving and analyzing mobile apps feature requests from online reviews*. in *2013 10th Working Conference on Mining Software Repositories (MSR)*. 2013.
8. Malgaonkar, S., A. Khan, and A. Vichare. *Mixed bilingual social media analytics: case study: Live Twitter data*. in *2017 International Conference on Advances in Computing, Communications and Informatics (ICACCI)*. 2017.
9. Berander, P. and A. Andrews, *Requirements Prioritization*, in *Engineering and Managing Software Requirements*, A. Aurum and C. Wohlin, Editors. 2005, Springer Berlin Heidelberg: Berlin, Heidelberg. p. 69-94.
10. Store, G.P. cited 2017; Available from: https://play.google.com/store?hl=en.
11. Store, A.A. cited 2017; Available from: https://www.apple.com/nz/ios/app-store/.
12. Keertipati, S., B.T.R. Savarimuthu, and S.A. Licorish, *Approaches for prioritizing feature improvements extracted from app reviews*, in *Proceedings of the 20th International Conference on Evaluation and Assessment in Software Engineering*. 2016, ACM: Limerick, Ireland. p. 1-6.
13. Licorish, S.A., B.T.R. Savarimuthu, and S. Keertipati, *Attributes that Predict which Features to Fix: Lessons for App Store Mining*, in *Proceedings of the 21st International Conference on Evaluation and Assessment in Software Engineering*. 2017, ACM: Karlskrona, Sweden. p. 108-117.
14. Karlsson, L., et al., *Requirements engineering challenges in market-driven software development – An interview study with practitioners*. Information and Software Technology, 2007. **49**(6): p. 588-604.
15. Bagnall, A.J., V.J. Rayward-Smith, and I.M. Whittley, *The next release problem*. Information and Software Technology, 2001. **43**(14): p. 883-890.
16. Laurent, P., J. Cleland-Huang, and C. Duan. *Towards Automated Requirements Triage*. in *15th IEEE International Requirements Engineering Conference (RE 2007)*. 2007.
17. Petersen, K., et al., *Systematic mapping studies in software engineering*, in *Proceedings of the 12th international conference on Evaluation and Assessment in Software Engineering*. 2008, BCS Learning & Development Ltd.: Italy. p. 68-77.
18. Sher, F., et al. *Requirements prioritization techniques and different aspects for prioritization a systematic literature review protocol*. in *2014 8th. Malaysian Software Engineering Conference (MySEC)*. 2014.